\documentstyle[preprint,aps]{revtex}
\begin{document}
\title{The $\pi \pi$ S-Wave in the 1 to 2 GeV Region from a
$\pi \pi$, $\bar{K}K$ and $\rho \rho$($\omega \omega$) Coupled Channel Model}
\author{W.M. Kloet}
\address{Department of Physics and Astronomy, Rutgers University,
PO Box 849, Piscataway, New Jersey 08855, USA}
\author{B. Loiseau}
\address{Division de Physique Th\'eorique \footnote[1]{Unit\'e de
Recherche des Universit\'es Paris 11 et Paris 6 Associ\'ee au CNRS},
Institut de Physique Nucl\'eaire, F-91406, Orsay CEDEX, France\\
and LPTPE, Universit\'e P. \& M. Curie, 4 Place Jussieu, 75252, Paris
CEDEX 05, France}
\date{\today}
\maketitle
\begin{abstract}
A simple $\pi \pi$, $\bar{K}K$, and $\rho \rho$($\omega \omega$) fully coupled
channel model  is proposed to predict the isoscalar S-wave phase shifts and
inelasticities for $\pi \pi$ scattering in the 1.0 to 2.0 GeV region. The
S-matrix is required to exhibit poles corresponding to the established
isoscalar J$^{\pi}$ = 0$^+$ resonances f$_0$(975), f$_0$(1400), and
f$_0$(1710).  A dominant feature of the experimental $\pi \pi$ inelasticity is
the clear opening of the $\bar{K}K$ channel near 1 GeV, and the opening of
another channel in the 1.4 - 1.5 GeV region. The success of our model in
predicting this observed  dramatic energy dependence  indicates that the effect
of multi-pion channels is adequately described by the $\pi \pi$ coupling to the
$\bar{K}K$ channel, the $\rho \rho$(4$\pi$) and $\omega \omega$(6$\pi$)
channels.
\end{abstract}
\pacs{ }

\narrowtext

\section{INTRODUCTION}

In heavy ion  collisions and in anti-nucleon physics  many pions are produced
in the final state.  Pion-pion scattering will therefore play an important role
in the final state interaction. Our knowledge of $\pi \pi$ scattering is
incomplete, in particular above $M_{\pi \pi} \approx$ 1~GeV.   The dynamics of
$\pi \pi$ scattering is often described   by effective meson-exchange in the
t-channel. However one may also consider the presence of many resonances in the
s-channel, that  distinguishes $\pi \pi$ scattering from, for example,
nucleon-nucleon  scattering. For energies larger than 1 GeV an additional
aspect is  its strong coupling to multi-pion channels, for example, four and
six pions.

Previous models for $\pi \pi$ scattering  have been applied from threshold to
1.4 GeV.  For example separable  potential models have been
considered~\cite{can1,can2,can3} in an approach in which the major focus was to
find the right combination of attraction and repulsion in the diagonal $\pi
\pi$ interaction, to describe the proper low energy  behavior of $\pi \pi$
S-wave phase, and to better understand the structure of the f$_0$(975). More
sophisticated one-meson exchange models~\cite{Loh,Mul,Pea}  have been used up
to 2 GeV, also within the framework of coupling $\pi \pi$ and $\bar{K} K$
channels.

We have concentrated our work in the energy region of 1.0 -- 2.0 GeV, as one of
our interests is the study of final state interaction in
$\bar{p}p~\rightarrow~\pi \pi$ annihilation~\cite{Bat}, which requires
knowledge of $\pi \pi$ scattering in the 2 GeV energy range. Relevant
experimental $\pi \pi$ phase shifts and inelasticities were taken from the
analysis  by Hyams et al~\cite{Hya}. This group has extracted a pion-pion
scattering amplitude in the $M_{\pi \pi} = $ 1.02 - 1.78 GeV region from pion
production in pion-nucleon scattering. Two of their solutions and the low
energy data from Ref.~\cite{Ros,Gra} are plotted in Fig. 1. Note  the rather
strong  energy dependence of the phase shift and the inelasticity.  We do not
know of any higher energy data, so a model reproducing the existing data is
needed for an extrapolation to 2 GeV. We have redone the $\pi \pi$ and
${\bar{K} K}$ coupled channel calculation of Refs.~\cite{can1,can2} up to 2
GeV. We confirm their calculation, but note that above 1.4 GeV the prediction
fails to reproduce the data as can be seen in Fig. 1. The dashed curve
represents set 2 of Table 1 of Ref.~\cite{can1} and the solid curve set 2  of
Table 2 of Ref.~\cite{can2}. The lack of agreement shows the necessity to
introduce more channels.

In view of the presence of an extensive set of resonances in $\pi \pi$
scattering, we have  opted in this paper for a separable potential model
similar to that of Refs.~\cite{can1,can2,can3}.
In the energy range of 1.0 - 2.0
GeV the dominant channels are $\pi \pi$, $\bar{K}K$, $\rho \rho$, and $\omega
\omega$. Within our model, as no spin effects have been taken into account,
the $\rho \rho$ and $\omega \omega$ channels can be described by a single
effective channel, e.g. $\rho \rho$. The existing S-wave  resonances that
couple strongly to the $\pi \pi$ channel are the f$_0$(975),  f$_0$(1400), and
f$_0$(1710). We impose the condition that the scattering  matrix should have
poles at the corresponding complex energies.

Section II describes the model and shows the analytic expression of the
S-matrix. Section III lists the parameter values for which the pole conditions
are satisfied and gives the corresponding phase shifts and inelasticities.
Discussion of the results follows in section IV.

\section{MODEL}

As in Ref.~\cite{can3} we consider three channels but, in contrast with  that
work, all channels are completely coupled giving a different expression for the
$\pi \pi$ S-matrix. Our method of derivation follows closely
that of Ref.~\cite{can1}. We describe briefly how to obtain the scattering
matrix. The three reduced Schr\"odinger equations take the form

\begin{equation}
-u^"_i(r) + \sum^3_{l=1} I_{il}(r) = k^2_i u_i(r).\label{schr}
\end{equation}
Here the relative momenta, k$_i$, in the i$^{th}$ channel, are related by the
kinematic condition

\begin{equation}
\frac{1}{4} M^2_{\pi \pi} = k^2_1+m^2_1 = k^2_2+m^2_2 = k^2_3+m^2_3.
\end{equation}
Furthermore

\begin{equation}
I_{il}(r) =  m_i r \int dr' r' V_{il}(r,r') u_l(r').
\end{equation}
Here the channels i=1,2,3 correspond to $\pi \pi$, $\bar{K} K$, and
$\rho \rho$ respectively. We used $m_1$=0.1396, $m_2$=0.4937 and $m_3$=0.7680
GeV. We ignore the width of the $\rho$ meson.

The interaction in this coupled three-channel model with separable  S-wave
potentials is described by  a symmetric real 3x3 matrix of the form

\begin{equation}
V_{ij}  (r,r') = \frac{^\lambda ij}{\sqrt{m_im_j}} ~ g_i (r) g_j (r').
\end{equation}
We choose the commonly used form factors

\begin{equation}
g_i(r) = e^{- \beta_i r} / r .  \label{ff}
\end{equation}
For the above potentials, solutions of the Eqs.\ \ref{schr} can be
written as

\begin{equation}
u_i(r) = c_i \left[ e^{ik_ir} + a_i e^{-ik_ir} - (1+a_i) e^{-\beta_ir}\right].
\end{equation}
For incoming pions in channel 1 a$_1$ is non-zero, a$_2$ = a$_3$ = 0 and the
S-matrix element for $\pi \pi \rightarrow \pi \pi$ is

\begin{equation}
S_{11} = -\frac{1} {a_1}.
\end{equation}
The integrals $I_{il}(r)$ can be done analytically and the coupled
Schr\"odinger equations lead to three algebraic equations with unknowns
$c_2/c_1$, $c_3/c_1$ and $a_1$. One obtains

\begin{equation}
S_{11} = D(-k_1,k_2,k_3) / D(k_1,k_2,k_3),
\end{equation}
where D is the Jost function

\begin{eqnarray}
D(k_1,k_2,k_3) = [( R_1 + \Lambda_1 ) (R_2 + \Lambda_2 ) (R_3 +
\Lambda_3 ) - \Lambda^{2}_{12} (R_3 + \Lambda_3 )
- \Lambda^{2}_{13}
(R_2 +\Lambda_2 ) \nonumber\\
- \Lambda^{2}_{23}  (R_1 +\Lambda_1 ) + 2
\Lambda_{12} \Lambda_{13} \Lambda_{23}]/(R_1 R_2 R_3).
\end{eqnarray}
Here, because of the special form factors of Eq.\ref{ff},

\begin{equation}
R_i  = ( 1 - i k_i / \beta_i)^2.
\end{equation}
The quantities $\Lambda_i = \lambda_{ii} / (2 \beta^{3}_{1})$, and
$\Lambda^{2}_{ij} = \lambda^{2}_{ij} / (4 \beta^{3}_{i}
\beta^{3}_{j})$ are renormalized dimensionless couplings.

We now require the S-matrix to have poles at the known J$^{\pi}$ = 0$^+$
resonances f$_0$(975), f$_0$(1400), and f$_0$(1710). There are three range
parameters $\beta _i$, and six couplings $\lambda_{ij}$. The three poles are
then  chosen at the  complex energies, (Re$M_{\pi \pi}$, Im$M_{\pi \pi}$),
respectively  (0.975, -0.0165), (1.400, -0.075), and (1.709, -0.350) GeV
providing six constraints on the nine parameters. The remaining freedom in the
parameter space is narrowed down by the requirement that all
off-diagonal couplings must be real.

\section{RESULTS}

We present the values of the nine parameters and the corresponding values of
the $\pi \pi$ phase shifts, $\delta$, and inelasticities, $\eta$, as a function
of total energy, $M_{\pi \pi}$, following the conventional parameterization of
the scattering matrix element S$_{11}$ = $\eta$ e$^{2i \delta}$. Several sets
of parameter values are given  in Table 1. All solutions for the parameters
share the feature that the interaction in the $\pi \pi$ channel is repulsive
and the interactions in the $\bar{K}K$ and $\rho \rho$ channels are mainly
attractive. The form factors in the $\pi\pi$ and $\rho\rho$ channels are
relatively long ranged, while for the $\bar{K}K$ channel the form factor can be
either short or long ranged.

The resulting $\delta$ and $\eta$, corresponding to the second (solid
line)
and fourth (dashed line)  sets of parameters of Table 1,
are given in Fig. 2.  As expected, the energy dependence of the
inelasticity is closely related to the opening of the $\bar{K}K$
and $\rho \rho$ channels (see Fig. 2.). This is also
clearly seen from the band of solutions for $\eta$ representing the entire
Table 1, shown in Fig. 3. The exact value of $\eta$ is very sensitive to the
model parameters, in particular $\beta_{3}$  and the corresponding  coupling
$\lambda_{13}$ of the $\pi \pi$ and $\rho \rho$ channels. The values of the
phase  shifts are far less sensitive to the choice of parameters.
Again this is seen from the spread of solutions for $\delta$ shown
in Fig. 3.  The phase shift is about 180 degrees at the opening of the
$\bar{K}K$ channel and approximately 400 degrees at the opening of the $\rho
\rho$ channel.

\section{DISCUSSION AND CONCLUSION}

Comparing with experimental analysis, we find the best agreement with the
solution set (--\,--\,--) and set (++\,--) of Ref.~\cite{Hya}. The  authors of
that work preferred set (--\,--\,--) at the time of their analysis.  We note
that the data point for $\delta$ in their (--\,--\,--) set at the
highest energy
could be increased by 180 degrees, which would be consistent with the sharp
rise in the (++\,--) data set. The global agreement of the predicted
inelasticity in this simple model for several parameter sets of Table
1 is remarkable (see Fig. 3.), even with  the wide scatter of the
experimental data.

The overall behaviour of the inelasticity is determined by the  opening of the
higher channels, i.e. the branch cut structure of the S-matrix. The $\rho
\rho$-branch cut begins at $M_{\pi\pi}$ = 2~m$_{\rho}$ = 1.536 GeV and that of
$\omega \omega$ at $M_{\pi \pi}$ = 2 m$_{\omega}$ =  1.564 GeV. Our model,
which contains a realistic $\rho \rho$ branch cut at $M_{\pi \pi}$ = 1.536 GeV,
shows that the inelasticity starts to decrease sharply at about 1.4 GeV.
The experimental data exhibit a very similar behaviour. The data  does not show
any sign of a specific 4$\pi$-branch cut near 0.558 GeV  or 6$\pi$-branch cut
near 0.837 GeV. This indicates that the production of four and six
pions occurs mainly via $\rho \rho$ and $\omega \omega$  formation.

The behaviour of the phase shift is insensitive to modest variation of the
ranges and couplings. However, $\pi \pi$ phase shifts can be changed by
modifying the position of the  resonances, i.e. the pole structure of the
S-matrix. For example, increasing the mass of the third resonance to 1.900 GeV
brings the phase shift into somewhat better agreement with the data,
and at the same time preserves the structure of the inelasticity. This
is illustrated in Fig. 4.

In conclusion, we have constructed a three-channel model
which gives a reasonable description of the
S-wave  $\pi \pi$ scattering in the 1.0 - 1.7 GeV region. This has been
accomplished by requiring the S-matrix to have poles at the three known
isoscalar 0$^+$ resonances as well as realistic branch cuts at the
$\bar K K$ and
$\rho \rho (\omega \omega)$ thresholds. Multi-pion channels are described as
effective $\rho \rho$ and $\omega \omega$ channels.  The model can be used to
predict the S-wave $\pi \pi$ interaction in the 2 GeV region which can play an
important role in the study of the recently well measured  $\bar{p}p
\rightarrow \pi \pi$ process.

\acknowledgements

One of the authors (W.M.K.) is grateful to the Division de
Physique Th\'eorique, Institut de Physique Nucl\'eaire at Orsay for
its hospitality during his stay, when this work was initiated.
 Both authors thank W.R. Gibbs and R.Vinh Mau for helpful discussions and
also acknowledge the Theory Group, T-5, of Los Alamos National Laboratory.

\begin{figure}
\caption[]{$\pi \pi$ phase shifts (in degrees) and inelasticities
from various experimental analyses. Data points (+) are from set  (--\,--\,--)
and data points (x) are from set (++\,--) of Ref. \cite{Hya}.  Solid points
below 0.6 GeV are data from Ref.~\cite{Ros} and earlier data referred to there.
Solid points above 0.6 Gev are from Ref~\cite{Gra}. The curves represent $\pi
\pi$ phase shifts and inelasticities from previous two-channel
models~\cite{can1,can2}.  Dashed curve has one term in the $\pi\pi$ and the
$\bar{K}K$ channel, while the solid curve has an additional attractive term in
the $\pi\pi$ channel.}
\end{figure}

\begin{figure}
\caption{$\pi \pi$ phase shifts and
inelasticities from our three-channel model.
Solid curve is the prediction for parameter set 2, and dashed curve
for set 4, of Table 1. Data are as in Fig.~1. }
\end{figure}

\begin{figure}
\caption{Spread of $\pi \pi$ phase-shift and
inelasticity solutions from three-channel model for all parameter sets of
Table 1.}
\end{figure}

\begin{figure}
\caption{Spread of $\pi \pi$ phase-shift and
inelasticity solutions
from three-channel model where the resonance at 1.709 GeV has been shifted
to 1.900 GeV.}
\end{figure}

\mediumtext

\begin{table}
\caption{Model parameters ($\beta$ in GeV).}
\begin{tabular}{ccccccccc}
$\beta _1$ & $\beta _2$& $\beta _3$ &$\Lambda _1$ & $\Lambda_2$&
$\Lambda_3$ & $\Lambda_{12}$ & $\Lambda_{13}$ & $\Lambda_{23}$\\
\tableline
0.175&  0.250&  0.225 &  17.80 &  0.588 & -5.10 &  22.60 &  5.18 &10.700\\
0.175&  0.250&  0.250 &  17.90 &  0.157 & -4.20 &  21.60 &  7.48 & 6.750\\
0.175&  0.250&  0.275 &  18.00 & -0.349 & -3.50 &  20.00 &  9.86 & 3.140\\
0.175&  0.250&  0.300 &  18.60 & -0.719 & -3.20 &  16.30 & 12.40 & 0.687\\
0.150&  0.400&  0.200 &  23.60 & -0.456 & -4.70 &  15.40 & 17.50 & 4.530\\
0.150&  0.400&  0.250 &  23.90 & -0.619 & -3.60 &  14.80 & 15.70 & 2.030\\
0.150&  0.400&  0.300 &  24.40 & -0.699 & -3.10 &  14.00 & 14.00 & 1.000\\
0.150&  0.400&  0.350 &  24.80 & -0.791 & -2.70 &  13.40 & 12.60 & 0.384\\
0.150&  0.400&  0.200 &  23.60 & -0.456 & -4.70 &  15.40 & 17.50 & 4.530\\
0.150&  0.700&  0.200 &  25.20 & -0.897 & -4.70 &   6.89 & 31.80 & 0.442\\
0.150&  1.000&  0.200 &  26.00 & -0.944 & -4.90 &   4.51 & 34.30 & 0.090\\
0.150&  1.500&  0.200 &  26.70 & -0.963 & -5.10 &   2.91 & 35.00 & 0.017\\
\end{tabular}
\end{table}

\end{document}